# Excess Mechanical Loss Associated with Dielectric Mirror Coatings on Test Masses in Interferometric Gravitational Wave Detectors

D. R. M. Crooks, P. Sneddon, G. Cagnoli, J. Hough
Dept of Physics and Astronomy
University of Glasgow, Glasgow G12 8QQ, UK

S. Rowan, M. M. Fejer, E. Gustafson, R. Route
Edward L. Ginzton Laboratory
Stanford University, Stanford, California 94305-4085, USA

N. Nakagawa
Centre for Nondestructive Evaluation, Institute for Physical Research and Technology,
Iowa State University, Ames, Iowa 50011, USA

D. Coyne
LIGO Project, California Institute of Technology, Pasadena, California 91125

G. M. Harry
LIGO Project, Massachusetts Institute of Technology, Room NW17-161,
175 Albany Street, Cambridge, Massachusetts 02139, USA
Department of Physics, Syracuse University, Syracuse, New York 13244-1130

A. M. Gretarsson
Department of Physics, Syracuse University, Syracuse, New York 13244-1130

**Abstract**

Interferometric gravitational wave detectors use mirrors whose substrates are formed from materials of low intrinsic mechanical dissipation. The two most likely choices for the test masses in future advanced detectors are fused silica or sapphire [1]. These test masses must be coated to form mirrors, highly reflecting at 1064nm. We have measured the excess mechanical losses associated with adding dielectric coatings to substrates of fused silica and calculate the effect of the excess loss on the thermal noise in an advanced interferometer.

## 1. Introduction

Interferometric gravitational wave detectors use laser interferometry to sense the position of test masses coated as mirrors, suspended as pendulums and highly isolated from external disturbances. Across part of the frequency range of interest for gravitational wave detection, thermal noise from the test masses and their suspensions, forms a limit to achievable detector sensitivity [2,3]. To minimize the thermal noise from the test masses, substrate materials of low intrinsic loss are desirable. All the detectors currently under construction use fused silica as the substrate material, due to a combination of its properties including relatively low mechanical loss, availability in suitably sized pieces, and ease of polish. To benefit from the low intrinsic dissipation of the substrates all sources of excess mechanical loss associated with using the substrates as suspended mirrors should be minimised [4,5]. The multi-layer dielectric coatings which are applied to each substrate to form mirror coatings are a potential source of excess dissipation, and we present here measurements of the excess dissipation associated with typical coatings. To assess the level of dissipation the loss factors of samples of fused silica with dielectric coatings were measured. Similar measurements on substrates of different geometry and with a coating of different constituents were simultaneously carried out by a subset of the authors at Syracuse University and are published in an accompanying paper. From these measurements it is possible to estimate values for the mechanical loss associated with the dielectric coatings. Using the work of Nakagawa et al, [6,7] the effect of this loss on the expected sensitivity of advanced gravitational wave detectors may then be calculated.

For advanced detectors sapphire is also under consideration as a substrate material, since it can have an intrinsic dissipation even lower than fused silica. Mechanical losses associated with adding coatings to sapphire substrates are a subject of ongoing research.

**2. Experimental measurements on fused silica samples**

Two coated fused silica samples were studied, each being a right circular cylinder of 0.127 m diameter and 0.103 m height. The first sample was made from Corning 7980 fused silica (grade 0C) and the second from Corning 7940 fused silica (3G). Each sample was polished by General Optics Inc. [8], having faces super-polished to sub-angstrom roughness levels and barrels with an inspection polish. Dielectric coatings were applied to each face of the cylinders by the same company. One face of each sample had a mirror coating designed to be highly reflecting at 1064 nm, and an anti-reflection coat (at 1064 nm) was applied to each rear face. From our analysis we believe that the high reflective coating on the 7980 mass consists of approximately forty-three alternating quarter wavelength layers of aluminium oxide and tantalum pentoxide, with a geometrical thickness of 6.3 µm. For the 7940 mass we believe there were approximately 59 layers with a geometrical thickness of 8.6 µm.

We make the assumption that the loss in the coating is proportional to the total coating thickness and that the loss is homogenous throughout the coating, thus we assume that the anti-reflection coating (which has significantly fewer layers ~2) does not add appreciably to the total loss. In the accompanying paper describing results obtained by colleagues at Syracuse University the possibility of the coating loss being different depending on whether the coating is deformed perpendicular to the surface, parallel to the surface or in shear, is examined. There will be further discussion of this later.

The loss factors of 7 modes of each sample were measured. The experimental arrangement used to measure the loss factors is shown in Figure 1.

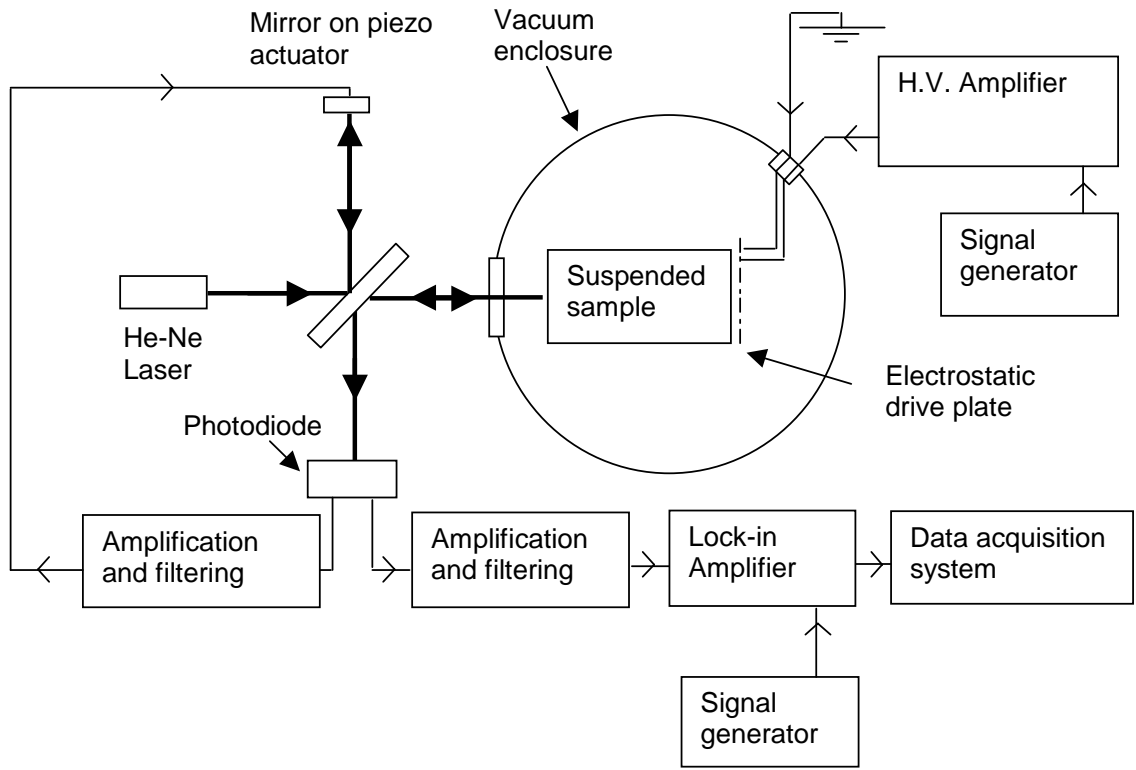

Figure 1 Experimental arrangement for measuring the loss factors of suspended test mass samples

The samples were suspended using a single sling of lightly greased tungsten wire of 150 μm diameter, with the upper ends of the wires held inside a steel clamp, mounted on a rigid tripod structure.  This structure was mounted inside a vacuum tank, evacuated to approximately $10^{-5}$ mb. The resonant modes of each sample were excited using an electrostatic drive plate mounted behind the samples, and the amplitude of the induced motion of the front face of each sample was sensed using a Michelson interferometer illuminated by light from a helium neon laser. The mirror in the reference arm of the Michelson was mounted on a piezoelectric transducer which, using a signal from the photodiode sensing the interference signal, was locked at low frequency to the pendulum motion of the mass on its suspension. The high frequency motion of the front face of the mass, sampled by the laser beam, was obtained from the signal from the photodiode, at frequencies well above the unity gain point of the servo loop. By measuring the rate of decay of the amplitude of the resonant motion of the test masses, the loss factors of the modes of the samples were obtained.

It should be noted that the samples were suspended multiple times, with the lengths of the suspension wires being varied each time, and the lowest measured loss factors for each mode used in our analysis. The reason for this is that previous experiments [9,10] have shown that the measured loss factors may be too large if the frequency of the resonant mode of the test mass happens to coincide with a frequency of a resonant mode of the suspensions wires.

## 3. Results

The lowest measured loss factors for 7 modes of the coated 7940 and 7980 fused silica masses are shown in Table 1.

|  | | Corning 7980 | | Corning 7940 | |
|---|---|---|---|---|---|
| Mode (*) | Modeled Frequency (Hz) | Measured Frequency (Hz) | Measured loss (x $10^{-7}$) | Measured Frequency (Hz) | Measured loss (x $10^{-7}$) |
| 1. Bending (8, n=1) | 22401 | 22105 | 1.37 +/- 0.04 | 22361 | 1.6+/-0.01 |
| 2. Asymmetric Drum (1, n=0) | 23238 | 22977 | 1.16 +/- 0.02 | 23004 | 1.23+/-0.05 |
| 3. Fundamental (1, n=2) | 24671 | 25378 | 0.65 +/- 0.01 | 25404 | 0.5+/-0.02 |
| 4. Clover 4 (16, n=2) | 25490 | 26176 | 1.61 +/- 0.03 | 26193 | 1.89+/-0.04 |
| 5. Symmetric Drum (4, n=0) | 27723 | 28388 | 3.1 +/- 0.12 | 28395 | 3.6+/-0.29 |
| 6. Expansion | 31397 | 31710 | 1.09 +/- 0.01 | 31731 | 1.01+/-0.01 |
| 7. 2nd Asymmetric Drum (3, n=0) | 35133 | 36045 | 0.86 +/- 0.01 | 36072 | 0.94+/-0.03 |

Table 1: Experimental losses for Corning 7980 and 7940 silica test masses, with dielectric coating as described in section 2. (Errors show 1 σ level). *Numbering denotes the symmetry classification of the modes following G. McMahon [11]

It can be seen that there is a considerable variation in the level of measured loss factor between the modes, with some modes showing a loss factor as much as 4 times higher than others and we postulate that this difference is due to losses associated with the coating. Previous measurements of an uncoated 7980 test mass of the same dimensions, but different inclusion class, showed a much smaller variation of the loss factor between equivalent modes. The loss factors varied from 0.87 x $10^{-7}$ to 1 x $10^{-7}$, lower than the loss factors for most of the modes of the coated masses. The spread in measured loss factors was within 14%. Thus we believe that suspension losses are insignificant and that the substantial variation in loss measurements seen for the modes of the coated samples is predominantly due to the effects of the dielectric coating. In each case a small amount of the coating had spilled over on to the barrel of the mass during the coating process. As will be seen later this is an important effect.

Assuming that all other losses have been reduced to a negligible level the total measured loss may be expressed as the sum of the intrinsic loss of the substrate material plus any loss associated with having added a coating to the substrate. In the general case of coating both on the face and on the barrel:

$$\phi(\omega_0)_{coated} = \frac{E_{substrate}}{E_{total}} \phi(\omega_0)_{substrate} + \frac{E_{coating\,on\,face}}{E_{total}} \phi(\omega_0)_{coating\,on\,face} + \frac{E_{barrel\,coating}}{E_{total}} \phi(\omega_0)_{eff} \quad (1)$$

and, assuming $E_{coating\,on\,face} << E_{substrate}$ and $E_{barrel\,coating} << E_{substrate}$

$$\phi(\omega_0)_{coated} \approx \phi(\omega_0)_{substrate} + \frac{E_{coating\ on\ face}}{E_{substrate}} \phi(\omega_0)_{coating\ on\ face} + \frac{E_{barrel\ coating}}{E_{substrate}} \phi(\omega_0)_{eff} \quad (2)$$

Where $E_{coating\ on\ face}/E_{substrate}$ is the fraction of the energy of the mode stored in the coating compared with the substrate and $E_{barrel\ coating}/E_{substrate}$ is the fraction of the energy of the mode stored in the barrel compared with the substrate.

$\phi(\omega_0)_{eff}$ represents the effective loss due to the coating material on the barrel of the sample and it is assumed at this stage that the distribution of the coating on the barrel is essentially even and of the same thickness as the coating on the faces. This assumption will be discussed further at a later stage.

Finite element analysis [12] was used to model the displacement of the test masses for each of the modes under study - these mode shapes are shown in Appendix 1. The relevant energy ratios for the faces and for the barrels of the different modes were then calculated. It was assumed that the coating followed the contours of the silica mass below. Using typical bulk values for Young's modulus and Poisson's ratio for aluminium oxide and tantalum pentoxide, an equivalent value for each property was obtained for the multilayer coating as described in [13]. These values were used in calculating the energy associated with the layers of coating, whereas the relevant values for fused silica were used for the substrate. The values used are as in Table 2.

| Material | Young's Modulus (Pa) | Poisson's Ratio |
|---|---|---|
| Aluminium Oxide | 3.6 x $10^{11}$ [15] | 0.27 [14] |
| Tantalum Pentoxide | 1.4 x $10^{11}$ [13] | 0.23 [13] |
| Calculated Multilayer [13] | 2.6 x $10^{11}$ | 0.26 |
| Fused Silica | 7.2 x $10^{10}$ [14] | 0.17 [14] |

Table 2: Material properties for coatings and substrate

In carrying out this calculation, care was exercised as the ratios obtained depended on the coarseness of the mesh used in the FE analysis. In practice the ratios were computed for a range of grid dimensions and the results extrapolated to the case of an infinitely fine mesh.

The convergence of an energy ratio as a function of the number of nodes used in the FE model for a typical mode is shown in Figure 2 and the convergent energy ratios for the different modes are shown in Figure 3.

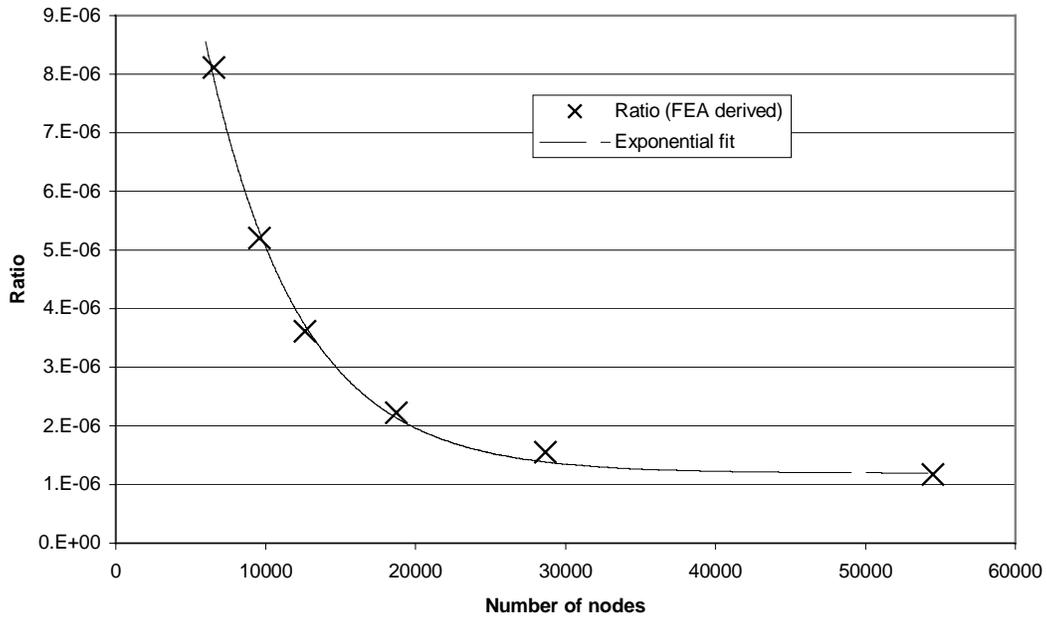

Figure 2: Variation in the computed ratio of energy stored in the dielectric mirror coating on a silica substrate per micron of coating thickness to the total energy stored in the substrate as the number of nodes in the FE model is increased. These data are for the asymmetric drum mode. The convergent value for this ratio is $1.19 \times 10^{-6}$.

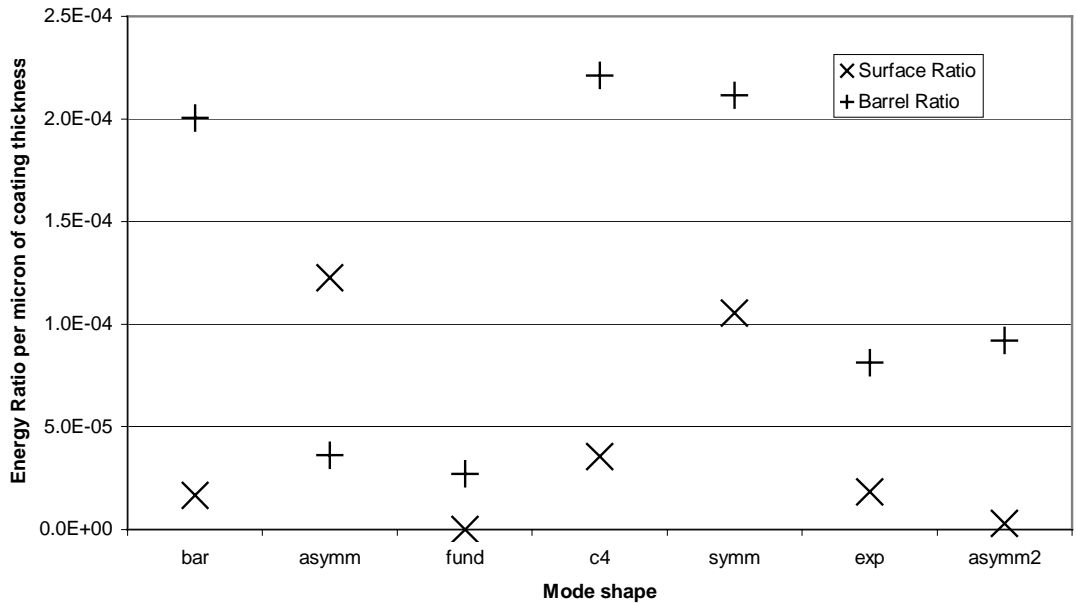

Figure 3: Ratios of energy stored per micron thickness of dielectric coating to total energy in substrate. Points X refer to the front surface and points + refer to the curved barrel.

## 3.1 Regression Analysis

As an initial approach it was assumed in equation 2 that $\phi_{eff} = 0$ (i.e. mechanical losses associated with the coating overspray on the barrel of the sample were of a level which was not significant for these measurements). Then the variation of $\phi_{coated}$ against front surface energy ratio was examined for the 7940 and 7980 test masses.

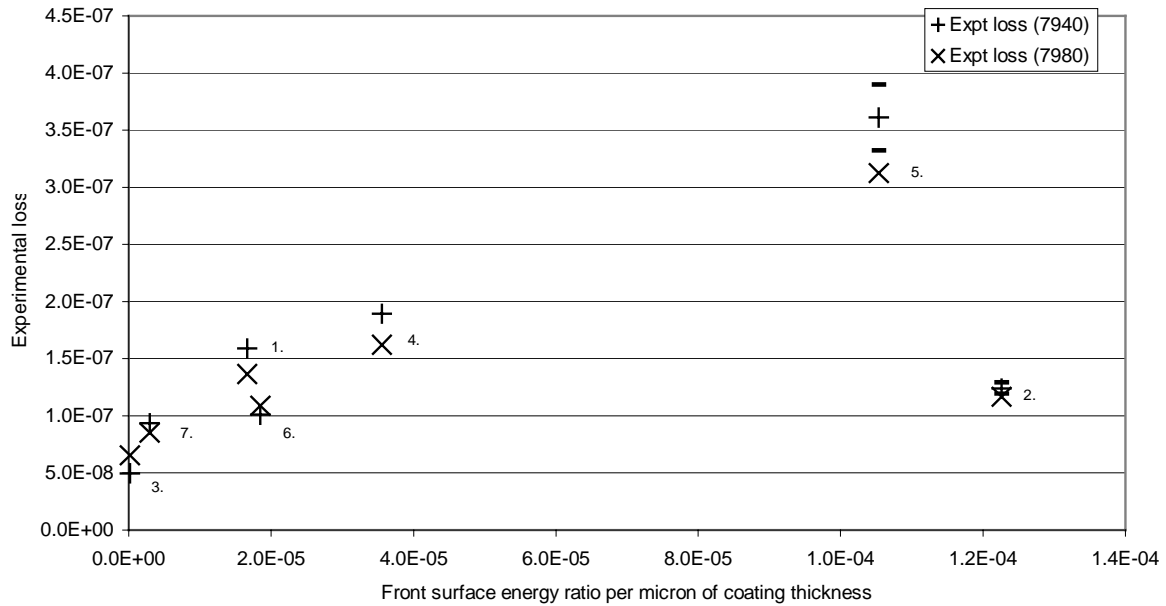

Figure 4: Variation of measured loss factor $\phi_{coated}$ against front surface energy ratio per micron thickness of coating for each mode of the 7940 and 7980 silica test masses. The experimental errors are highest for the two points for each mass at the right hand side of the figure and the size of these errors (1 $\sigma$ level) is indicated by the bars on the points for the 7940 mass. The numbering of the points refers to the mode numbering in Table 2.

It can be seen in Figure 4 that if one point in each case is excluded from the data, then for both data sets there is an excellent linear fit of coated loss to the front surface energy ratio. However the point which is excluded - that for the asymmetric drum mode - is one where the confidence in its value is particularly high since it is unlikely that experimental measurements will result in losses which are significantly too low. Thus it seems that the simple analysis is not adequate and a fuller analysis involving losses due to the coating on the barrels of the masses was carried out. In this case a multivariable linear regression algorithm for three parameters was adopted and it was no longer possible to use the simple graphical representation of Figure 4. The fits were now demonstrated by graphs of Experimental Loss against Predicted Loss for both the 7940 and the 7980 masses as in Figures 5 and 6.

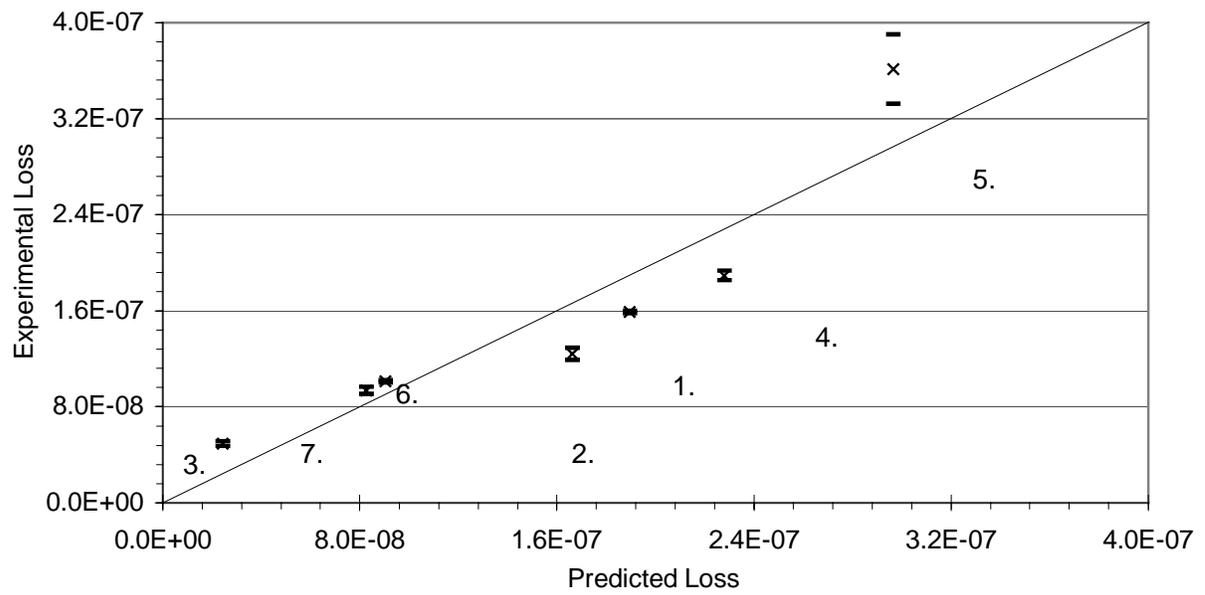

(a)

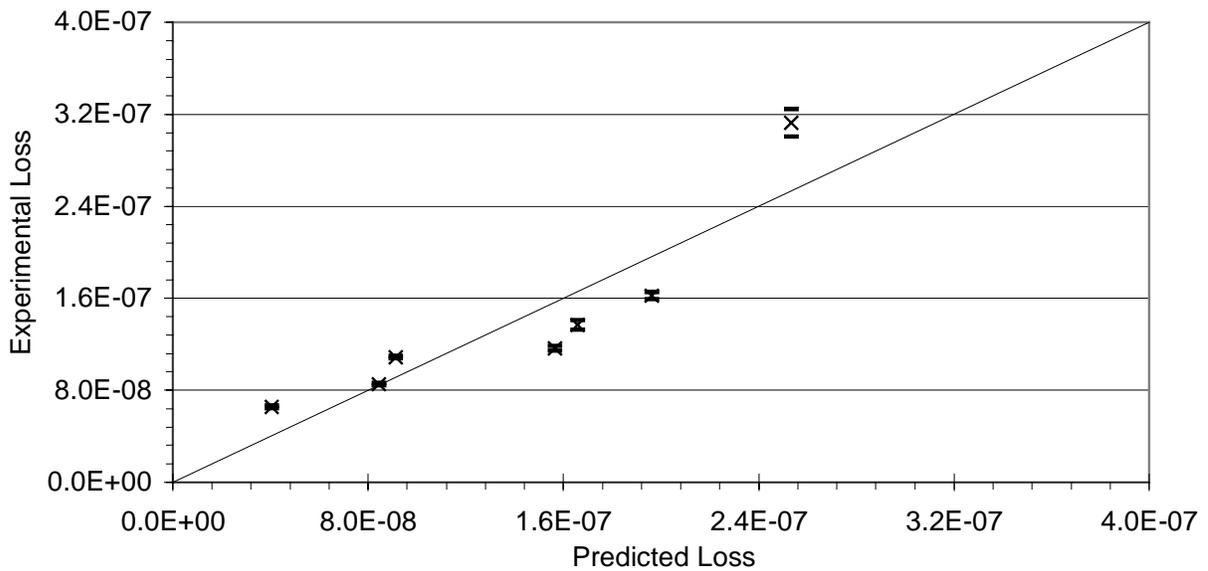

(b)

Figure 5: Comparison of the experimental loss with that predicted using a 3 parameter multiregression analysis for a) the 7940 and b) the 7980 mass. Mode numbers are shown on (a) only. The line y = x is drawn for information in each case

As can be seen the fit is likely to be good in both cases provided the point corresponding to the symmetric drum mode is ignored. This is clear from Figure 6 which shows the same results as above with the symmetric drum removed.

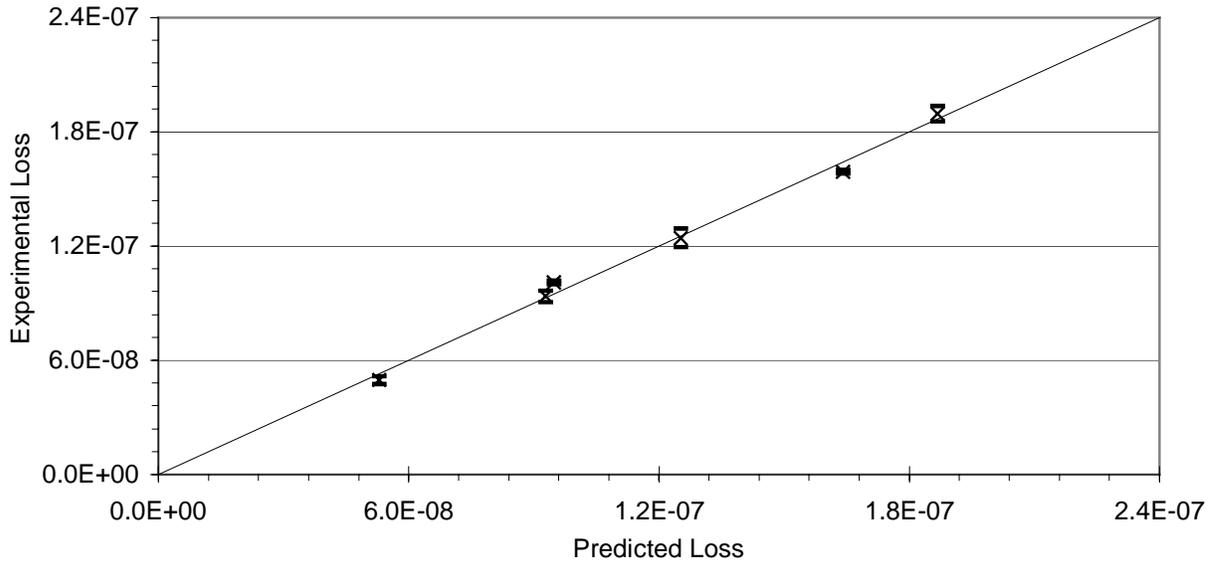

(a)

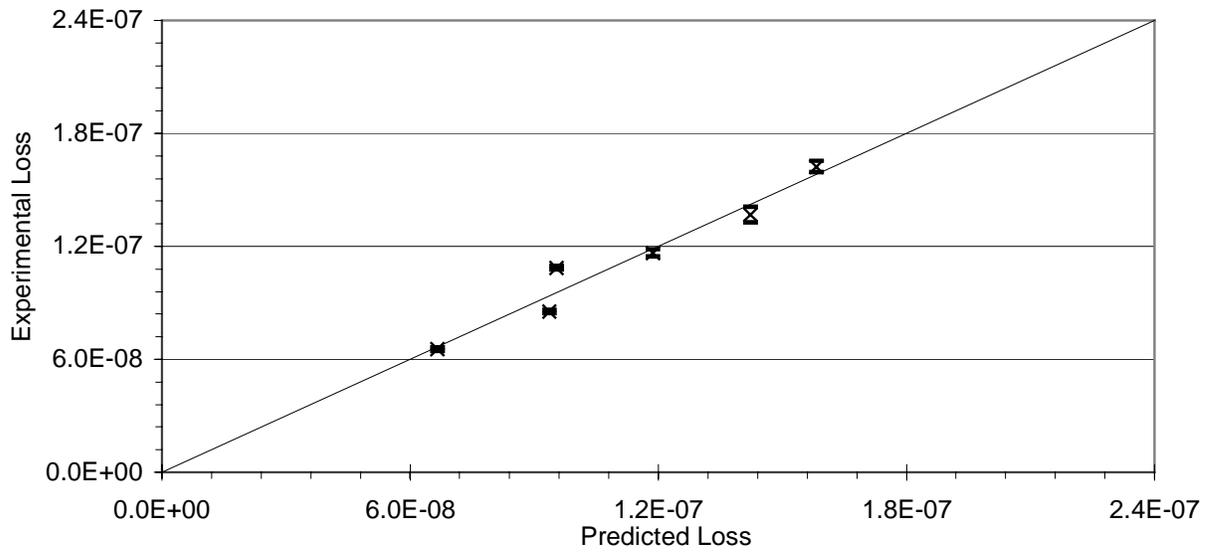

(b)

Figure 6: Comparison of the experimental loss with that predicted using a 3 parameter multiregression analysis for a) the 7940 and b) the 7980 mass, this time with the symmetric drum mode excluded. The line y = x is drawn for information.

The loss for the symmetric drum mode is significantly higher than that which is consistent with the best fit. Since this is the case for both the 7940 and 7980 masses it appears to be a real effect. In case this result is an artifact of the assumption of even coating thickness on the barrel a number of possibilities were investigated. For example situations where the coating effects on the barrel were mainly at one end or mainly in the middle were investigated. However no improvements were observed for the fitting of the data either with or without the symmetric drum mode included. Also, the possibility of the result being due to differing losses being associated with shear and bulk strain was studied. Analysis shows that the ratio of shear to bulk strain is broadly similar for all the modes except for the clover-4 mode (mode number 4) where shear dominates by a large factor. There is no sign that the loss for this clover mode is significantly different from that of the others suggesting that our assumption of a homogeneous loss is well founded. We also checked that the kinetic energy of the symmetric drum mode at the breakaway points of the suspending wire was not particularly high compared with that of the other modes and thus it seems unlikely that frictional loss could explain the observations. It has to be assumed that there is some un-modelled loss associated with the shape of the symmetric drum mode, perhaps due to edge effects at the chamfers between the coated mirror faces and the barrels.

The multivariable regression analysis, with the symmetric drum mode excluded, yields values for the loss parameters as follows:

For the Corning 7940 mass:
- $\phi(\omega_0)_{substrate} = (3.7 \pm 0.5) \times 10^{-8}$
- $\phi(\omega_0)_{coating} = (6.4 \pm 0.6) \times 10^{-5}$
- $\phi(\omega_0)_{eff} = (6.9 \pm 0.4) \times 10^{-5}$

For the Corning 7980 mass:
- $\phi(\omega_0)_{substrate} = (5.6 \pm 0.9) \times 10^{-8}$
- $\phi(\omega_0)_{coating} = (6.3 \pm 1.6) \times 10^{-5}$
- $\phi(\omega_0)_{eff} = (6.3 \pm 0.9) \times 10^{-5}$

Table 3: Fitted values of bulk, coating and barrel losses for the 7940 and 7980 masses. Errors are calculated from the regression analysis at the 1 σ level.

The values obtained for $\phi(\omega_0)_{substrate}$ are comparable with the best results for bulk samples of these types of fused silica [16].

**4. Significance of Coating Results for Advanced Detectors**

Until relatively recently the method of predicting the thermal noise in the test masses of gravitational wave detectors, at frequencies well below the first internal resonant modes of the mirrors, involved a normal mode expansion of the acoustic modes of the mirrors. The expected thermal noise was calculated by adding incoherently, with suitable weighting factors, the noise in the tails of the resonances of the test masses. Alternative approaches to the problem have been developed by Levin [17], Nakagawa et al [6], Bondu et al [18] and Liu and Thorne [19]. Levin has pointed out that in cases where the spatial distribution of the loss in a mirror is non-uniform, using a modal expansion approach may lead to an incorrect estimation of the thermal noise.
Nakagawa et al [6] have developed a formalism where the fluctuation dissipation theorem can be expressed in terms of the static Green's function to calculate the distribution of the thermal motion on the front face of a semi-infinite mirror. This work has been extended to allow estimation of the effect of having a coated layer of finite thickness and different material properties on the front face of the mass. It is calculated [7] that the spectral density of the thermal noise, $S_\phi^{coated}(f)$, is increased, over that of the uncoated mass, $S_\phi^{substrate}(f)$, by the factor shown in equation (3)

$$\frac{S_\phi^{coated}(f)}{S_\phi^{substrate}(f)} = \left\{1 + \frac{2}{\sqrt{\pi}} \frac{(1-2\sigma)}{(1-\sigma)} \frac{\phi_{coating}}{\phi_{substrate}} \left(\frac{d}{w}\right)\right\} \tag{3}$$

where *d* is the coating thickness, w is the radius of the laser beam interrogating the test mass, and $\sigma$ represents the Poisson's Ratio of the substrate material of Young's modulus *Y*. This assumes that the substrate and coating have the same mechanical properties but different loss factors.

More fully, allowing different values for $\sigma$ and *Y* in the coating and substrate the ratio then becomes [20].

$$\left\{1 + \frac{1}{\sqrt{\pi}} \frac{(1+\sigma_{coating})}{(1-\sigma_{substrate}^2)(1-\sigma_{coating})} \frac{\phi_{coating}}{\phi_{substrate}} \frac{Y_{susbstrate}}{Y_{coating}} \left[(1-2\sigma_{coating}) + (1-2\sigma_{substrate})^2 \frac{(1+\sigma_{susbtrate})^2}{(1+\sigma_{coating})^2} \left(\frac{Y_{coating}}{Y_{substrate}}\right)^2\right] \left(\frac{d}{w}\right)\right\}$$

Using the moduli of elasticity and Poisson's ratio for the coating material and the substrate listed in Table 2, a laser beam whose spot has radius $5.5 \times 10^{-2}$ m on the coating surface [21], the mean of the bulk and coating losses for the 7940 and 7980 masses listed in Table 3 and using the mean of the coating thicknesses for the 7980 and 7940 masses the predicted power spectral density of the thermal noise is found to be increased by a factor (1 + 0.27). This means that the amplitude spectral density of the noise is increased by 1.13 over that which would be predicted without evaluating the effect of the mechanical loss of the coating. It should be noted that we have assumed that the losses in the coating and the substrate are structural in nature, i.e. they have the same value at low frequency as they have at the mode frequencies where they were measured. This seems reasonable as our calculations suggest that thermoelastic damping – often non-structural in nature - is at a lower level than the losses we have measured.

## 5. Conclusion

Experiments suggest that the effect of argon-sputtered dielectric coatings on the level of thermal noise associated with silica test masses in advanced interferometric gravitational wave detectors will be significant. Experiments on sapphire substrates are under way and preliminary analysis suggests a similar damping effect by the coating. There is a clear need for a series of experiments to be carried out, in which coating parameters are systematically varied as this will allow the source of the coating losses to be investigated.

**Acknowledgements**

The authors would like to thank their colleagues in the University of Glasgow, the University of Hanover, the Max Planck Institut für Quantenoptik Garching, Stanford University and Syracuse University for their interest in this work. The authors would also like to thank Helena Armandula for useful discussions on coating. They are also grateful to PPARC in the UK, NSF in the US (grant PHY-99-00793) and their Universities for financial support.

**Appendix 1**

**Pictures of the test mass modes**

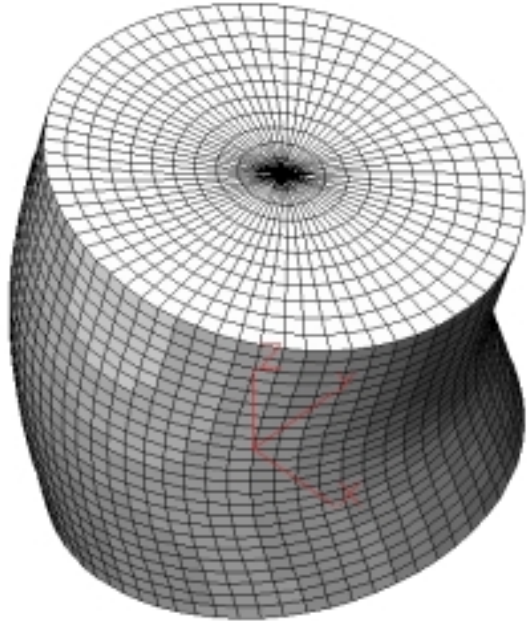

"Bar" mode

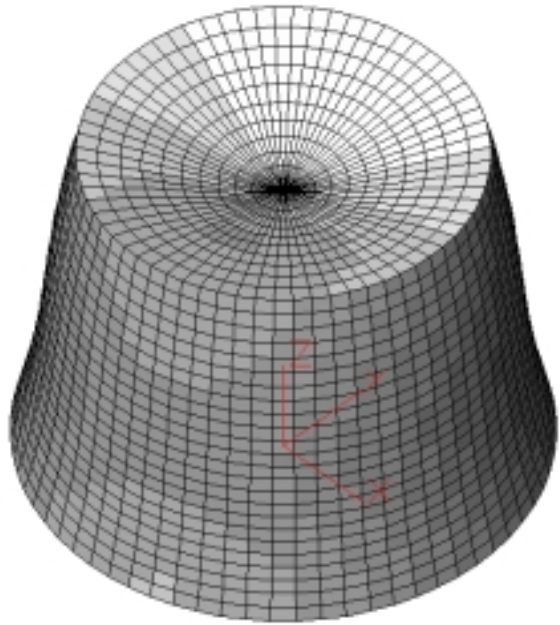

"Asymmetric drum" mode

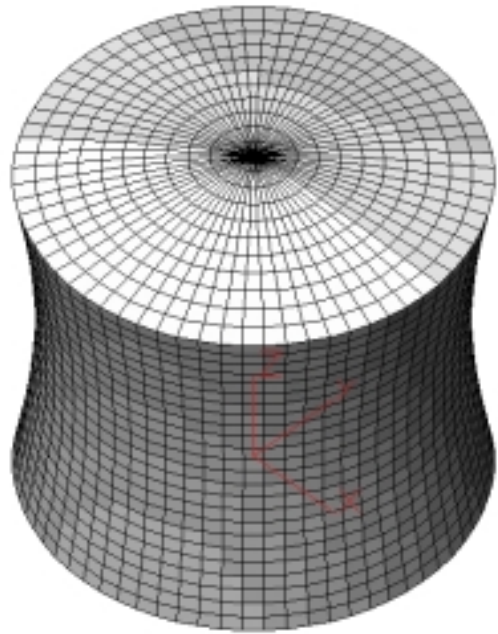

"Fundamental" mode

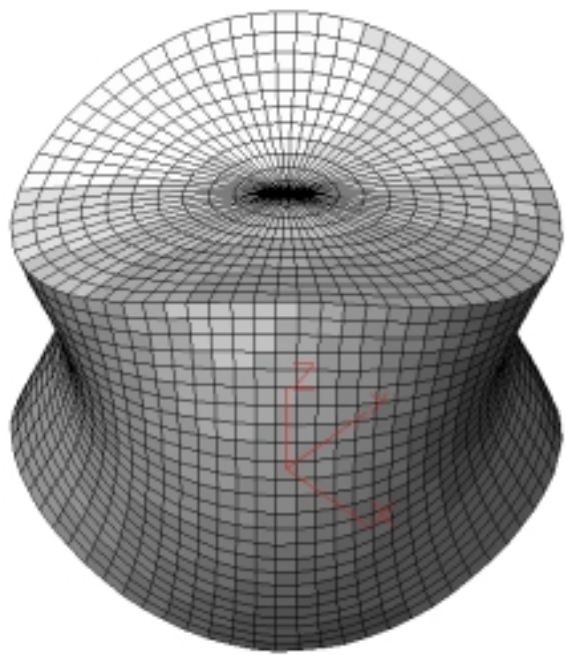

"Clover 4" mode

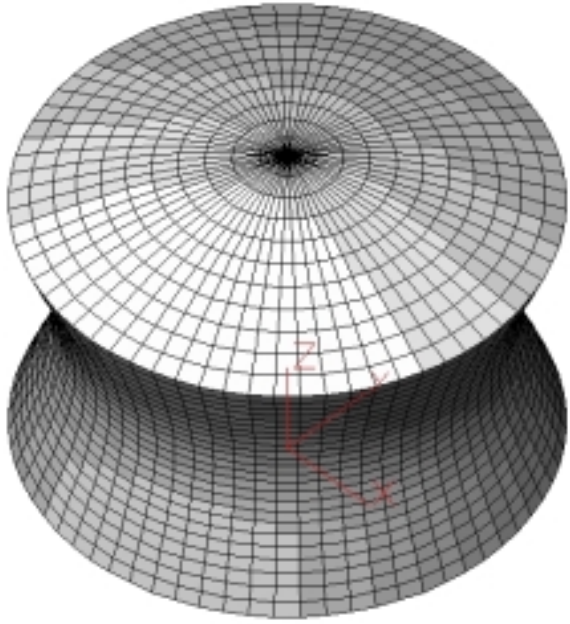

"Symmetric drum" mode

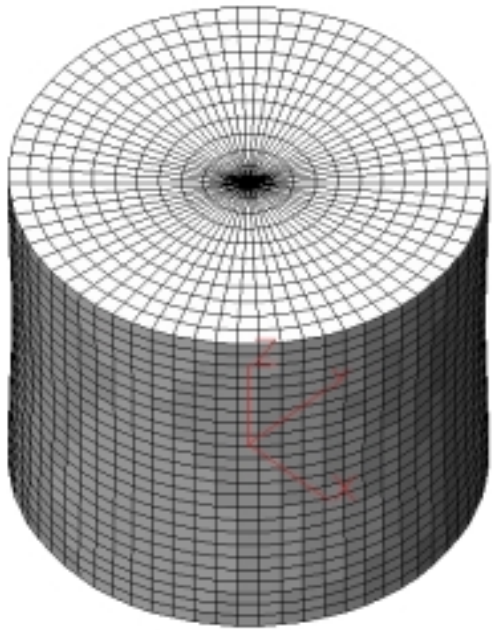

"Expansion" mode

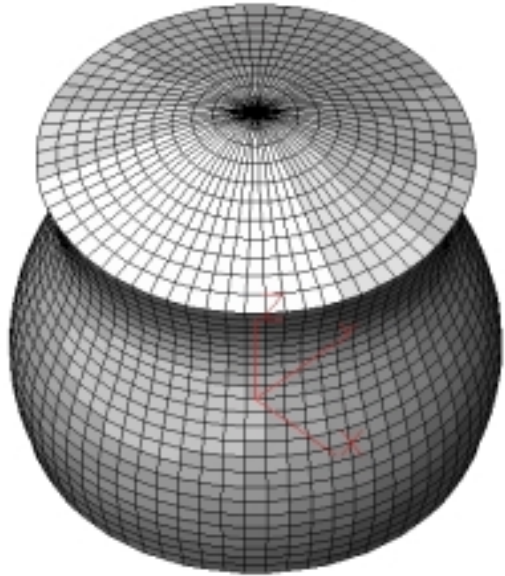

"Second asymmetric drum" mode